\documentclass[twocolumn,english,prl,superscriptaddress]{revtex4}
\usepackage[T1]{fontenc}
\usepackage[latin9]{inputenc}
\usepackage{amsmath}
\usepackage{graphicx}
\usepackage{amssymb}
\usepackage{esint}

\makeatletter
\@ifundefined{textcolor}{}
{%
 \definecolor{BLACK}{gray}{0}
 \definecolor{WHITE}{gray}{1}
 \definecolor{RED}{rgb}{1,0,0}
 \definecolor{GREEN}{rgb}{0,1,0}
 \definecolor{BLUE}{rgb}{0,0,1}
 \definecolor{CYAN}{cmyk}{1,0,0,0}
 \definecolor{MAGENTA}{cmyk}{0,1,0,0}
 \definecolor{YELLOW}{cmyk}{0,0,1,0}
 }

\makeatother

\makeatother

\usepackage{babel}

\begin{document}

\title{Spectrum of an electron spin coupled to an unpolarized bath of nuclear
spins}

\author{Oleksandr Tsyplyatyev}

\affiliation{Department of Physics, University of Basel, Klingelbergstrasse 82,
CH-4056 Basel, Switzerland}

\affiliation{Department of Physics \& Astronomy, University of Sheffield, Sheffield
S3 7RH, United Kingdom}

\author{Daniel Loss}

\affiliation{Department of Physics, University of Basel, Klingelbergstrasse 82,
CH-4056 Basel, Switzerland}
\begin{abstract}
The main source of decoherence for an electron spin confined to a
quantum dot is the hyperfine interaction with nuclear spins. To analyze
this process theoretically we diagonalize the central spin Hamiltonian
in the high magnetic $B$-field limit. Then we project the eigenstates
onto an unpolarized state of the nuclear bath and find that the resulting
density of states has Gaussian tails. The level spacing of the nuclear
sublevels is exponentially small in the middle of each of the two
electron Zeeman levels but increases super-exponentially away from
the center. This suggests to select states from the wings of the distribution
when the system is projected on a single eigenstate by a measurement
to reduce the noise of the nuclear spin bath. This theory is valid
when the external magnetic field is larger than a typical Overhauser
field at high nuclear spin temperature. 
\end{abstract}

\date{\today}

\maketitle
Spin dynamics in semiconductor nanostructures has recently become
a topic of great interest due to the possibility of using the spin
degree of freedom instead of charge in electronic circuits \cite{DasSarmaHanson}
and equally important due to the proposal of using electron spin in
a semiconductor quantum dot as a fundamental building block of the
quantum computing device \cite{LossDiVincenzo}. GaAs quantum dots
are the main candidates in practical realizations of these proposals
due to the well developed manufacturing technology. However, unavoidable
inhomogeneous hyperfine interaction of electron spin with many nuclear
spins of the host crystal acts as a noisy environment that is the
main source of dephasing for the electron spin at low temperature
when relaxation due to the phonons is ineffective.

The limit of fully polarized nuclear spin bath was analyzed exactly
in \cite{KhaetskiiLossGlazman}, including spectral properties. However,
it is rather hard to achieve a significant polarization dynamically,
and thermodynamic polarization, requiring sub-milli Kelvin temperatures
\cite{SBL}, is still out of reach for semiconductors. Currently,
a more promising route is to actively reduce the distribution width
of the nuclear Overhauser field by projective measurements \cite{CoishLoss,KlauserCoishLoss,StepanenkoBurkardImamoglu}.
This has been partially achieved in experiments leading to significantly
longer decoherence times \cite{VandersypenLocking,Bayer,BluhmYacoby}.
To further optimize projective measurement techniques it is essential
to gain a better understanding of the spectral properties of the unpolarized
system which, so far, have only been understood qualitatively.

\begin{figure}[t]
 \center{\includegraphics[width=1\columnwidth]{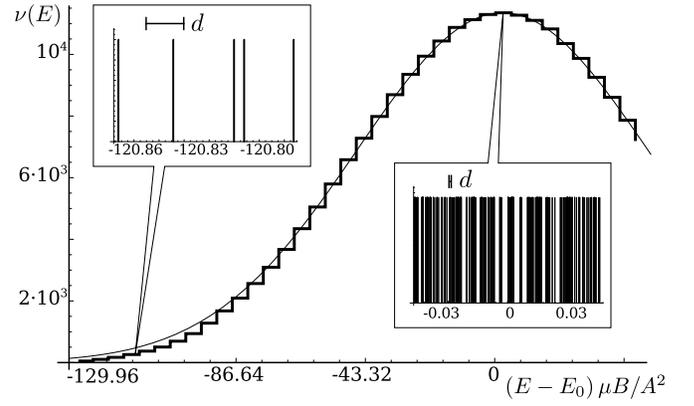}}

\caption{\label{fig:fig}Numerical evaluation of $\nu\left(E\right)$ using
Eq. (\ref{eq:E}) on a course scale - thick line and Eq. (\ref{eq:PDOS})
- thin line ($S_{j}=1$, $r_{0}=8$, $N=18$, a fixed external $B$-field),
$A$ is a maximum Overhauser field, $E_{0}$ is a shift from Eq. (5).
Insets show $\nu\left(E\right)$ on a fine scale in the middle of
the upper electron Zeeman line and at a finite detuning, the average
level spacing $d$ was evaluated using Eq. (\ref{eq:level_spacing}).}

\end{figure}

In this paper we diagonalize the central spin Hamiltonian for a quantum
dot in the high magnetic $B$-field limit using a $1/B$-expansion.
Projecting the eigenstates on an unpolarized state of the nuclear
spin bath we find that their density has Gaussian tails. Correspondingly
the level spacing of the nuclear spin sublevels, which is exponentially
small with the radius of the quantum dot in the middle of the two
electron Zeeman levels, becomes super-exponentially large with detuning
away from the center, see Fig. \ref{fig:fig}. This suggests using
a finite detuning from the bare electron Zeeman energy when one eliminates
the effect of the nuclei by the projective measurement technique \cite{VandersypenLocking,Bayer,BluhmYacoby,KlauserCoishLoss,StepanenkoBurkardImamoglu,CoishLoss}.%

Our theory is applicable when the external magnetic field $B$ is
larger than a typical Overhauser field at high nuclear spin temperature
due to fluctuations $B_{\textrm{fluc}}=A\sqrt{S/\tilde{N}}/\mu$,
where $A/\mu$ is the maximum Overhauser field, $\tilde{N}$ is the
number of nuclei under the electron envelope wave function, and $S$
is a number of degenerate hyperfine couplings. At low field $B<B_{\textrm{fluc}}$
the spectrum can be obtained by a numerical solution of the Richardson
equations \cite{Gaudin} where the $1/B$-expansion of the present
paper can be used as a benchmark for complex numerical procedures.

The spin of an electron in a quantum dot couples to nuclear spins
in the presence of an external $B$-field as

\begin{equation}
H=\mu BS_{0}^{z}+\sum_{j=1}^{N}A_{j}\mathbf{S}_{0}\cdot\mathbf{S}_{j},\label{eq:H0}\end{equation}
where $\mu=g\mu_{B}$ is the electron magneton (in the following we
neglect the nuclear Zeeman splitting), $S_{0}^{z},S_{0}^{\pm}=S_{0}^{x}\pm iS_{0}^{y}$
are electron spin-1/2 operators and $\mathbf{S}_{j}$ ($j\geq1$)
are spin operators of nuclear shell $j$ with the maximum angular
momentum $S_{j}\geq1/2$ constructed out of $2S_{j}$ nuclei of spin-1/2
which have the same hyperfine coupling to the electron spin, $\mathbf{S}_{j}=\sum_{i=1}^{2S_{j}}\mathbf{I}_{ji}$,
where $i$ labels individual nuclei within the shell, $\mathbf{I}_{ji}$
are nuclear spin-1/2 operators, and $N$ is the number of nuclear
shells. Assuming harmonic confinement of the electron in all spatial
directions the couplings are $A_{j}=A_{0}\exp\left(-r_{j}^{2}/r_{0}^{2}\right)$,
where $A_{0}$ is the coupling in the middle of the quantum dot, and
$r_{0}$ and $r_{j}$ are spatial size of the quantum dot and radius
of jth shell in units of the lattice parameter.

In 1D only two nuclei have the same coupling ignoring the isotope
effects and assuming equidistant lattice sites $r_{j}=j$, thus the
maximum total angular momentum is $S_{j}=S=1$. In 2D degeneracy of
the couplings gives $S_{j}=S=4$ but the radii of the sequential shells
are not equidistant because the number of nuclei grows linearly away
from the center. We thus model the system as a set of concentric nuclear
shells, $r_{j}=r+4m/\left(\pi r\right)$ and also change the summation
indices in Eq. (1), $\sum_{j=1}^{N}\rightarrow\sum_{r=1,m=1}^{N,\pi r/4}$
\cite{supplementary}. In 3D the degeneracy is larger than in 2D,
$S_{j}=S=12$, and the number of the nuclei grows quadratically away
from the center, $r_{j}=r+6m/\left(\pi r^{2}\right)$, $\sum_{j=1}^{N}\rightarrow\sum_{r=1,m=1}^{N,\pi r^{2}/6}$.

This model conserves the number of excitations $\left[H,J_{z}\right]=0$,
where $J_{z}=\sum_{j=0}^{N}S_{j}^{z}$, and the total angular momentum
of each nuclear shell $\left[H,\mathbf{S}_{j}^{2}\right]=0$. All
of them also commute with each other, $\left[J_{z},\mathbf{S}_{j}^{2}\right]=0$
and $\left[\mathbf{S}_{i}^{2},\mathbf{S}_{j}^{2}\right]=0$. Thus
the Hilbert space is partitioned into a set of disconnected subspaces
labeled by the following quantum numbers: $n$ is an eigenvalue of
$J_{z}$ and $l_{j}$ \cite{degeneracy} correspond to $\mathbf{S}_{j}^{2}$,
$\mathbf{S}_{j}^{2}\left|\Psi\right\rangle =l_{j}(j_{j}+1)\left|\Psi\right\rangle $.
The latter becomes trivial when all of the nuclear spins have different
couplings as for spin-1/2 operators $\mathbf{S}_{j}^{2}=3/4$ is a
number but is nontrivial when $S_{j}>1/2$.

The diagonalization in each subspace can be performed using degenerate
perturbation theory when the $B$-field is large. Splitting the Hamiltonian
into the unperturbed part $H_{0}=\mu BS_{0}^{z}$ and a perturbation
$V=\sum_{j}A_{j}\mathbf{S}_{0}\cdot\mathbf{S}_{j}$ defines two electron
Zeeman levels, $E=\pm\mu B/2$ but leaves the nuclear spin sublevels
hugely degenerate in the zeroth-order approximation. The latter degeneracy
has to be lifted via a diagonalization of the perturbation $V$.

In the basis of eigenstates of $J_{z}$, $\left|\Psi\right\rangle =\left|\pm,\left\{ l_{j},k_{j}\right\} \right\rangle $,
$\left\langle \Psi|\Psi\right\rangle =1$, $V$ is a diagonal matrix
within both of the electron spin subspaces where the spin-flip part
of $V$ that couples opposite electron levels can be neglected when
the external field is very large. Here $\pm$ refers to the {}``up''
and {}``down'' electron Zeeman levels and $k_{j}$ are the numbers
of nuclear spin excitations on each shell such that the quantum number
$n=\left(1\pm1\right)/2+\sum_{j=1}^{N}k_{j}$. The second order correction
to the eigenenergies are due to the spin-flip part of $V$. Using
the matrix elements of $V$ in the basis of eigenstates of $J_{z}$
we obtain

\begin{multline}
E=\pm\frac{\mu B}{2}\pm\sum_{j=1}^{N}\bigg[\frac{A_{j}\left(-l_{j}+k_{j}\right)}{2}\\
+\frac{A_{j}^{2}\left(2l_{j}-k_{j}+\frac{1\mp1}{2}\right)\left(k_{j}+\frac{1\pm1}{2}\right)}{4\mu B}\bigg],\label{eq:E}\end{multline}
where the energy denominator in the last term was also expanded up
to the leading order in $1/\mu B$. Including the first order corrections
to the eigenfunctions we get \begin{equation}
\left|\Psi\right\rangle =\left|\pm,\left\{ l_{j},k_{j}\right\} \right\rangle \pm\sum_{m=1}^{N}\frac{A_{m}}{\mu B}S_{m}^{\pm}\left|\mp,\left\{ l_{j},k_{j}\right\} \right\rangle .\label{eq:eigenfunctions_highB}\end{equation}

The large magnetic field expansion has different conditions of applicability
for the eigenenergies Eq. (\ref{eq:E}) and the eigenstates Eq. (\ref{eq:eigenfunctions_highB})
in the subspaces of unpolarized nuclear spins $k_{j}\approx l_{j}$.
The subleading terms in Eq. (\ref{eq:E}) are small in all subspaces
when $B\gg B_{\textrm{fluc}}$ where $B_{\textrm{fluc}}=\sqrt{\sum_{j=1}^{N}A_{j}^{2}S_{j}^{2}}/\mu.$
But the next (second) subleading correction to Eq. (\ref{eq:eigenfunctions_highB})
is small only when $B\gg B_{\textrm{max}}$ where $B_{\textrm{max}}=r_{0}^{2}A_{0}/2\mu$
in 1D and 2D ($B_{\max}=r_{0}^{3}A_{0}/\sqrt{8e}\mu$ in 3D) \cite{supplementary}
is a much larger field than $B_{\textrm{fluc}}$. The latter signals
that the choice of the eigenfunctions, $\left|\Psi\right\rangle =\left|\pm,\left\{ l_{j},k_{j}\right\} \right\rangle $,
is a poor zeroth order approximation in the intermediate field regime,
$B_{\textrm{fluc}}\ll B\ll B_{\textrm{max}}$. The correct approximation
can be identified by merging the inner nuclear shells with different
couplings up to the radius $\tilde{r}=r_{0}\sqrt{\ln\left(B_{\textrm{max}}/B\right)}$
(in units of the lattice parameter) in 1D and 2D ($\tilde{r}=r_{0}\left(1+\sqrt{\ln\left(B_{\textrm{max}}/B\right)}\right)/\sqrt{2}$
in 3D) \cite{supplementary} into a single shell with the same coupling
$A_{0}$. Then, diagonalizing $H+V'$, where $V'=\sum_{j:r_{j}\leq\tilde{r}}\left(A_{1}-A_{j}\right)\mathbf{S}_{0}\cdot\mathbf{S}_{j}$,
when $B_{\textrm{fluc}}\ll B\ll B_{\textrm{max}}$ instead of the
original model $H$ we obtain the same result as in Eqs. (\ref{eq:E},
\ref{eq:eigenfunctions_highB}) but a different definition of nuclear
shells $\tilde{S}_{j}$, where the first element is $\tilde{S}_{1}=\sum_{j:r_{j}\leq\tilde{r}}S_{j}$,
the middle elements are $\tilde{S_{j}}=0$ for $1<r_{j}\leq\tilde{r}$,
and the outer elements, $r_{j}>\tilde{r}$, are $\tilde{S}_{j}=S_{j}$.

In 2D and 3D the parameter $B_{\max}$ is proportional to the measurable
maximum Overhauser field $A=\sum_{j=1}^{N}S_{j}A_{j}$, $A/\mu$ is
of the order of a few Tesla \cite{Safronov&vonKlitzing}, with the
numerical factor $\pi^{-1}$ and $\left(2\pi^{3}e\right)^{-1}$. In
1D, $B_{\max}=\tilde{N}A/\left(\pi\mu\right)$ is much larger than
$A$, here $\tilde{N}=\sum_{j=1}^{N}2S_{j}A_{j}/A_{0}$. The parameter
$B_{\textrm{fluc}}=A\sqrt{S/\tilde{N}}/\mu$ scales with the number
of nuclei under the electron envelop function in all dimensions.

In terms of density of states the bare electron level acquires a finite
smearing due to coupling to many degrees of freedom of unpolarized
nuclear spins. When the quantum dot is empty the nuclei at different
lattice sites are uncorrelated. After an electron, say with spin {}``up'',
populates the quantum dot, the state of the combined system $\left|\Psi_{0}\right\rangle =S_{0}^{+}\prod_{\left\{ j,i\right\} }I_{ji}^{+}\left|\Downarrow\right\rangle $
is not an eigenstate of the Hamiltonian Eq. (\ref{eq:H0}), where
$\left\{ j,i\right\} $ labels a subset of nuclear lattice sites and
$\left|\Downarrow\right\rangle $ is the all spins down (including
the central spin) state. We analyze the distribution of the eigenenergies
Eq. (\ref{eq:E}) using a projected density of states $\nu\left(E\right)=\sum_{\left\{ l_{j},k_{j}\right\} }P\left(\left\{ l_{j},k_{j}\right\} \right)\delta\left(E-E\left(\left\{ l_{j},k_{j}\right\} \right)\right)$,
where $P\left(\left\{ l_{j},k_{j}\right\} \right)=1$ when $\left\langle \Psi_{0}|\left\{ l_{j},k_{j}\right\} \right\rangle \neq0$
and $P\left(\left\{ l_{j},k_{j}\right\} \right)=0$ when $\left\langle \Psi_{0}|\left\{ l_{j},k_{j}\right\} \right\rangle =0$.
Here the $\sum_{\left\{ l_{j},k_{j}\right\} }$ runs over all subspaces
and all eigenstates within each subspace. Note that for any shell with $S_j>1$ the complete set of the eigenstates includes $l_j$  with multiplicities greater than one \cite{degeneracy}. Only one of each $l_j$ is kept since these multiplicities do not change $P\left(\left\{ l_{j},k_{j}\right\} \right)$. We calculate the overlaps
matrix elements only in the leading $1/\mu B$-order as the probability
of measuring other eigenstates coming from subleading orders is at
least as small as $A_{j}/\mu B$.

By representing the delta function as $\delta\left(x\right)=\int d\lambda e^{\imath x\lambda}/\left(2\pi\right)$,
the Fourier transform of $\nu\left(E\right)$ can be written as a
product of sums over each nuclear spin shell \begin{multline}
\nu\left(\lambda\right)=\sum_{\left\{ l_{j},k_{j}\right\} }P\left(\left\{ l_{j},k_{j}\right\} \right)e^{-i\lambda E\left(l_{j},k_{j}\right)}\\
=\prod_{j=1}^{N}e^{-\frac{i\lambda\left(p_{j}A_{j}-\mu B\right)}{2}}\sum_{k=p_{j}\left(1+\textrm{sgn}p_{j}\right)}^{p_{j}+\tilde{S}_{j}}e^{-\frac{i\lambda A_{j}^{2}\left(k-2p_{j}\right)\left(k+1\right)}{4\mu B}},\label{eq:Fourier_nu}\end{multline}
where $p_{j}=\left\langle \Psi_{0}|S_{j}^{z}|\Psi_{0}\right\rangle $,
$\left|p_{j}\right|\leq l_{j}$, are polarizations of the shells given
by the state of the system $\left|\Psi_{0}\right\rangle $.

Assuming that each shell is unpolarized $p_{j}\ll\tilde{S}_{j}$ and
$\tilde{S}_{j}\gg1$, the sum within a shell can be calculated as
an integral, $I_{j}\left(\lambda\right)=\int_{0}^{\tilde{S}j}dke^{-ixk\left(k+1\right)}=\sqrt{\pi}e^{ix/4}\left[\textrm{erf}\left(\left(1+2\tilde{S}_{j}\right)\sqrt{ix}/2\right)-\textrm{erf}\left(\sqrt{ix}/2\right)\right]/\left(2\sqrt{ix}\right)$,
$x=\lambda A_{j}^{2}/\left(4\mu B\right)$, which is an oscillating
function of $\lambda$. Then the product of the oscillating functions
can be approximated in the large-$N$ limit by turning it into an
exponential of a sum of logarithms, $\prod_{j=1}^{N}I_{j}\left(\lambda\right)=I_{1}\left(\lambda\right)\exp\left(\sum_{j:r_{j}>\tilde{r}}^{N}\log I_{j}\left(\lambda\right)\right)$,
and by expanding the exponent in $\lambda$, $\sum_{j:r_{j}>\tilde{r}}^{N}\log I_{j}\left(\lambda\right)\approx\sum_{j:r_{j}>\tilde{r}}^{N}[\log S_{j}-i\left(S_{j}/2+S_{j}^{2}/3\right)\lambda A_{j}^{2}/\left(4\mu B\right)-(S_{j}^{2}/24+S_{j}^{3}/12+2S_{j}^{4}/45)\lambda^{2}A_{j}^{4}/\left(16\mu^{2}B^{2}\right)]$.

In 1D $I_{j}\left(\lambda\right)$ can not be calculated as an integral
since the degeneracy of the hyperfine couplings is two but the explicit
evaluation of the sum of only two terms within each shell and the
small-$\lambda$ expansion yields a similar expression, $\sum_{j:r_{j}>\tilde{r}}^{N}\log I_{j}\left(\lambda\right)\approx\sum_{j:r_{j}>\tilde{r}}^{N}[\log2-i\lambda A_{j}^{2}/\left(4\mu B\right)-\lambda^{2}A_{j}^{4}/\left(\sqrt{2}4\mu B\right)^{2}]$.
Strictly speaking, the small-$\lambda$ expansion is good when $\lambda\ll16\mu B/A_{\tilde{r}}^{2}$
but the resulting Gaussian is also quite a good approximation for
a large $\lambda$ since the original product of many oscillating
functions is zero due to random phases of $I_{j}\left(\lambda\right)$
when $\lambda\geq4\mu B/A_{\tilde{r}}^{2}$, provided that the couplings
$A_{j}$ have a non regular distribution.

By evaluating the inverse Fourier transform $\nu\left(E\right)=\int d\lambda\nu\left(\lambda\right)\exp\left(-iE\lambda\right)$
in the limit $B\gg B_{\max}$ we obtain \begin{equation}
\nu\left(E\right)=\frac{\tilde{S}_{1}\prod_{j:r_{j}>\tilde{r}}^{N}S_{j}}{\sqrt{\pi}\sigma}\exp\left[-\frac{\left(E-E_{0}\right)^{2}}{\sigma^{2}}\right],\label{eq:PDOS}\end{equation}
where $E_{0}=\sum_{j=1}^{N}p_{j}A_{j}/2-\mu B/2$ is a shift of the
bare electron level that depends on the momentary state of the nuclei
and a finite linewidth $\sigma=\sqrt{\sum_{j:r_{j}>\tilde{r}}^{N}\left(S_{j}^{2}/96+S_{j}^{3}/48+S_{j}^{4}/90\right)A_{j}^{4}}/\left(\mu B\right)\simeq\mu B_{\textrm{fluc}}^{2}/\left(\sqrt{\tilde{N}}B\right)$
that is common for all unpolarized nuclear states. In the intermediate
regime $B_{\textrm{fluc}}\ll B\ll B_{\max}$ Eq. (\ref{eq:PDOS})
is valid when $E\ge\tilde{S}_{1}^{2}A_{1}^{2}/\left(4\mu B\right)$.
The contribution of the inner shells can be approximated as $I_{1}\left(\lambda\right)=\tilde{S}_{1}$
when, due to the fast oscillating exponential, the main contribution
to the inverse Fourier transform comes from $\lambda\le4\mu B/\left(\tilde{S}_{1}A_{1}\right)^{2}$.

In 1D, the Gaussian result agrees precisely with the spectroscopically
measurable lineshape when $B\gg B_{\textrm{max}}.$ As the degeneracy
of hyperfine couplings is $2$ for all shells, all projections \cite{degeneracy}
are the overlap of the singlet (or triplet) and two nuclear spin states
which give $1/\sqrt{2}$ and the calculation of the lineshape gives
Eq. (\ref{eq:PDOS}). When the degeneracy is larger than $2$ the
two calculations are different. It is also worth noting that the state $\left|\Psi\left(0\right)\right\rangle $ is an eigenstate of the model Eq. (1) with $S_{j}=1/2$  in the high $B$-field $B\gg B_{\max}$.

Rediscretization of Eq. (\ref{eq:PDOS}) recovers the average level
spacing of the nuclear spin levels. From the definition of the density
of states, $d=1/\nu\left(E\right)$ is an energy range that contains
only one state. But, as the prefactor in $\nu\left(E\right)$ increases
to infinity when more and more outer shells are taken into account,
the level spacing becomes zero. On the other hand the coupling strengths
of the outer shells become super-exponentially small which make the
splitting of the inner shells' levels into sublevels due to the outer
shells very narrow. Thus, by selecting an effective number of the
significantly coupled nuclear shells $r_{j}<4r_{0}$, we find \begin{equation}
d\left(E\right)=d\left(E_{0}\right)\exp\left[\left(E-E_{0}\right)^{2}/\sigma^{2}\right],\label{eq:level_spacing}\end{equation}
where $d\left(E_{0}\right)=\sqrt{\pi}\sigma/\left(\tilde{S}_{1}\prod_{j:\tilde{r}<r_{j}<4r_{0}}S_{j}\right)$
is exponentially small, $d\left(E_{0}\right)\simeq\tilde{S}_{1}\exp\left(-\tilde{N}/S\right)$.
Thereby, $d\left(E_{0}\right)$ is a tiny level spacing in the middle
of the upper electron Zeeman line but $d\left(E\right)$ increases
super-exponentially at a finite detuning $E\neq E_{0}$ on a characteristic
energy scale $\sigma$ when $B\gg B_{\max}$ and $\tilde{S}_{1}^{2}A_{1}^{2}/\left(4\mu B\right)$
when $B_{\textrm{fluc}}\ll B\ll B_{\max}$.

There is also a finite temperature smearing. To average the hyperfine
shift $E_{0}$ over all possible nuclear spin configurations at a
high temperature, $\nu_{0}\left(E\right)=\sum_{\left\{ p_{j}\right\} }\delta\left(E-E_{0}\right)$,
we use the same approach as in the calculation of $\nu\left(E\right)$
and obtain the Gaussian distribution of levels with a width $\sigma_{0}=\sqrt{\sum_{j=1}^{N}S_{j}^{2}A_{j}^{2}/6}$
and an average level spacing $d_{0}=\exp\left[\left(E-\mu B/2\right)^{2}/\sigma_0^{2}\right]\tilde{S}_1\prod_{j:\tilde{r}<r_{j}<4r_{0}}2S_{j}/\left(\sqrt{\pi}\sigma_{0}\right)$.
This implies that if the nuclear spin state is not prepared in a specific
way but is a thermal state, there are two energy scales in a projective
measurement to narrow the nuclear spin bath \cite{BluhmYacoby,VandersypenLocking,Bayer}
in order to suppress fluctuations of the Overhauser field \cite{KlauserCoishLoss,CoishLoss,StepanenkoBurkardImamoglu}.
A measurement in the coarse resolution of $d_{0}$ will select a single
specific nuclear spin configuration suppressing only thermal fluctuations
and a measurement in the fine resolution of $d$ will project the
system on an eigenstate within a given nuclear bath state.

Using the eigenstates and the spectrum in Eqs. (\ref{eq:E}, \ref{eq:eigenfunctions_highB})
one can evaluate the time-dependent density matrix of the electron
with an unpolarized state of the nuclei, $\left|\Psi\left(0\right)\right\rangle =\left(1+S_{0}^{+}\right)\prod_{\left\{ j,i\right\} }I_{ji}^{+}\left|\Downarrow\right\rangle /\sqrt{2}$
such that $\left\langle \Psi\left(0\right)|J_{z}|\Psi\left(0\right)\right\rangle =0$,
as an initial condition. As a result the diagonal matrix elements
do not decay in time in the leading $1/\mu B$ order, $T_{1}=\infty$.
When the degeneracy of the hyperfine couplings is only $2$ (1D case
and $B\gg B_{\textrm{max}}$) the off-diagonal matrix elements have
a slow Gaussian envelop with decay time $T_{2}=1/\sigma$ on top of
the fast electron spin Rabi oscillations with frequency $\mu B$.
Note that one obtains the Gaussian decay assuming a phenomenological
model of a quasistatic ensemble of nuclear magnetic fields \cite{KlauserCoishLoss}.
At a high temperature, averaging over different $\left|\Psi\left(0\right)\right\rangle $,
one also obtains the Gaussian decay due to thermal fluctuations with
$T_{2}=1/\sigma_{0}$ \cite{SchliemannKhaetskiiLoss} which is much
faster than $1/\sigma$.

When the degeneracy of the hyperfine couplings is larger than $2$
(2D and 3D cases and $B_{\textrm{fluc}}\ll B\ll B_{\textrm{max}}$
in 1D) we establish a bound on the shortest decay time assuming that
all Clebsch-Gordon coefficients in the overlaps between the initial
state $\left|\Psi\left(0\right)\right\rangle $ and the eigenstates
Eq. (\ref{eq:eigenfunctions_highB}) are equal and neglecting degeneracies of $l_j$
\cite{degeneracy}. This simplification gives a Gaussian decay with
decay time $T_{2}=1/\sigma$. A more accurate calculation would give
a spectroscopic lineshape, see discussion after Eq. (\ref{eq:PDOS}),
which is narrower than the distribution of the eigenenergies thus
giving a longer decay time.

The eigenenergies Eq. (\ref{eq:E}) are a good benchmark for numerical
studies of Richardson equations \cite{Richardson}. The spectrum of
the model Eq. (1) can be found at arbitrary field and for any quantum
number $n$ by solving a set of coupled non-linear equations \cite{Gaudin},
\begin{equation}
\sum_{j=1}^{N}\frac{2l_{j}A_{j}/2}{E_{\nu}+A_{j}/2}+1-\frac{\mu B}{E_{\nu}}+\sum_{k=1\neq\nu}^{n}\frac{2E_{k}}{E_{\nu}-E_{k}}=0,\label{eq:Richardson_Eq}\end{equation}
as $E=\sum_{\nu=1}^{n}E_{\nu}+\sum_{j=1}^{N}l_{j}A_{j}/2-\mu B/2$.
At an infinitely large magnetic field solutions of these equations
are sets of numbers $E_{\nu}$ which are close either to $-A_{j}/2$
or $\mu B$. At a finite magnetic field a $1/B$-expansion of the
Eqs. (\ref{eq:Richardson_Eq}) at these values of $E_{\nu}$ recovers
the $1/B$-expansion in Eq. (\ref{eq:E}) and a $1/B$-expansion of
the Gaudin states \cite{Gaudin} recovers Eq. (\ref{eq:eigenfunctions_highB}).

In conclusion we have diagonalized the central spin Hamiltonian in
the high $B$-field limit. Projecting the eigenstates on an unpolarized
state of the nuclear bath we have shown that the level spacing of
the nuclear sublevels, which is exponentially small in the middle
of the bare electron level, becomes super-exponentially large with
detuning away from the middle. This suggests to select states from
the wings of the distribution when one attempts to eliminate the decohering
effect of the nuclei by projective measurement techniques. This theory
is valid when the external $B$-field is larger than typical Overhauser
fields.

We acknowledge support from the Swiss NF, NCCR Nanoscience Basel,
DARPA, IARPA, and EPSRC.

\appendix

\section*{Supplementary materials}

Here we provide more details on the calculations and approximations
of the main text. Figs. 1 and 2 illustrate construction of the shells
in the Hamiltonian Eq. (1) of the main text. Section A contians a
description of the zeroth order approximation to the eigenfunctions
in the intermediate field regime. In section B we solve explicitly
the Richardson equations in the large magnetic $B$-field limit.

\subsection{Intermediate field regime $B_{\textrm{fluc}}\ll B\ll B_{\textrm{max}}$}

The large magnetic field expansion has different conditions of applicability
for the eigenenergies,\begin{multline}
E=\pm\frac{\mu B}{2}\pm\sum_{j=1}^{N}\bigg[\frac{A_{j}\left(-l_{j}+k_{j}\right)}{2}\\
+\frac{A_{j}^{2}\left(2l_{j}-k_{j}+\frac{1\mp1}{2}\right)\left(k_{j}+\frac{1\pm1}{2}\right)}{4\mu B}\bigg],\label{eq:Ea}\end{multline}
 and the eigenstates, \begin{equation}
\left|\Psi\right\rangle =\left|\pm,\left\{ l_{j},k_{j}\right\} \right\rangle \pm\sum_{m=1}^{N}\frac{A_{m}}{\mu B}S_{m}^{\pm}\left|\mp,\left\{ l_{j},k_{j}\right\} \right\rangle ,\label{eq:eigenfunctions_highBa}\end{equation}
 in the subspaces of unpolarized nuclear spins $k_{j}\approx l_{j}$.
Comparison of the first and the third terms in Eq. (\ref{eq:Ea}) set
the limitation on external magnetic field as $B_{\textrm{fluc}}^{2}/B^{2}\ll1$
in all subspaces where

\begin{equation}
B_{\textrm{fluc}}=\frac{\sqrt{\sum_{j=1}^{N}A_{j}^{2}S_{j}^{2}}}{\mu}.\end{equation}
 The next order correction in Eq. (\ref{eq:Ea}) is small as $B_{\textrm{fluc}}^{3}/B^{3}$
compared to the leading term.

The second term of Eq. (\ref{eq:eigenfunctions_highBa}) is small when
$B\gg B_{\textrm{fluc}}$ but the next order correction is not. The
second order term in the flip-flop part of $V$ contains a denominator
which is a difference between unperturbed eigenenergies that belong
to the same electron spin level. This sets the limitation on the magnetic
field as \begin{equation}
\frac{A_{j}^{2}}{\left[2\mu B\left(A_{j}-A_{i}\right)\right]}\ll1\end{equation}
for all pairs $i\neq j$. Assuming that $r_{0}\gg1$, the most restrictive
condition comes from a pair of sequential shells at $r_{1}=1$ in
1D and 2D as $B/B_{\max}\ll1$, where \begin{equation}
B_{\textrm{max}}=\frac{r_{0}^{2}A_{0}}{2\mu}.\end{equation}
In 3D the most restrictive condition comes from a pair of sequential
shell at an intermediate radius $r_{j}=r_{0}/\sqrt{2}$ with \begin{equation}
B_{\textrm{max}}=\frac{r_{0}^{3}A_{0}}{\sqrt{8e}\mu},\end{equation}
where $e$ is the base of the natural logarithm. In all cases $B_{\textrm{max}}$ is a much larger field than $B_{\textrm{fluc}}$.

The divergence of perturbation series for the eigenstates Eq. (\ref{eq:eigenfunctions_highBa})
when $B_{\textrm{fluc}}\ll B\ll B_{\textrm{max}}$ does not mean that
a perturbation theory in $1/B$ is inapplicable at all but only signals
that the choice of the eigenfunctions, $\left|\Psi\right\rangle =\left(S_{0}^{+}\right)^{\left(1\pm1\right)/2}\prod_{j=1}^{N}\left(S_{j}^{+}\right)^{k_{j}}\left|\Downarrow\right\rangle $,
is a poor zeroth order approximation to split the nuclear spin sublevels
within one of the electron Zeeman levels. The correct approximation
can be identified by merging the inner nuclear shells (in 1D and 2D)
with different couplings up to the radius $\tilde{r}=r_{0}\sqrt{\ln\left(B_{\textrm{max}}/B\right)}$
into a single shell with the same coupling $A_{0}$ as the most restrictive
limitation originates from the shells at the middle of the electron
envelope function. In 3D the correct approximation can be identified
by merging the shells with intermediate radii between $\tilde{r}=r_{0}\left(1\pm\sqrt{\ln\left(B_{\textrm{max}}/B\right)}\right)/\sqrt{2}$.
The inner radius vanishes fast, when $B=B_{\textrm{max}}/e$, thus
we neglect it and merge all of the inner shells up to the radius $\tilde{r}=r_{0}\left(1+\sqrt{\ln\left(B_{\textrm{max}}/B\right)}\right)/\sqrt{2}$
in 3D as well as in 1D and 2D.

Then we diagonalize $H+V'$, where $V'=\sum_{j:r_{j}\leq\tilde{r}}\left(A_{1}-A_{j}\right)\mathbf{S}_{0}\cdot\mathbf{S}_{j}$,
instead of the original model $H$ from Eq. (1) of the main text by
repeating the same calculation as for $H$ and obtain Eqs. (\ref{eq:Ea},
\ref{eq:eigenfunctions_highBa}) with a different definition of nuclear
shells $\tilde{S}_{j}$, where the first element is $\tilde{S}_{1}=\sum_{j:r_{j}\leq\tilde{r}}S_{j}$,
the middle elements are $\tilde{S_{j}}=0$ for $1<r_{j}\leq\tilde{r}$,
and the outer elements, $r_{j}>\tilde{r}$, are $\tilde{S}_{j}=S_{j}$.
Corrections to this result due to $V'$ are small for $\left|r_{j}-\tilde{r}\right|\gg1$.
The specific form of the zeroth order eigenstates for $r_{j}\simeq\tilde{r}$
can only be found numerically and we neglect this crossover region
assuming a sharp transition between the two types of eigenstates.

\subsection{A solution for the Richardson equations}

The eigenenergies of the central spin Hamiltonian, \begin{equation}
H=\mu BS_{0}^{z}+\sum_{j=1}^{N}A_{j}\mathbf{S}_{0}\cdot\mathbf{S}_{j},\label{eq:H0}\end{equation}
at arbitrary external magnetic field in each subspace with a given
set of the quantum numbers $n$, $l_{j}$ can be found by solving
a set of Richardson equations \citep{Richardson}, \begin{equation}
\sum_{j=1}^{N}\frac{2l_{j}A_{j}/2}{E_{\nu}+A_{j}/2}+1-\frac{\mu B}{E_{\nu}}+\sum_{k=1\neq\nu}^{n}\frac{2E_{k}}{E_{\nu}-E_{k}}=0,\label{eq:Richardson_Eq}\end{equation}
 as \citep{Gaudin} \begin{equation}
E=\sum_{\nu=1}^{n}E_{\nu}+\sum_{j=1}^{N}l_{j}A_{j}/2-\mu B/2.\end{equation}
All sets of $E_{\nu}$s, which are the solution of the above set of
equations, also uniquely define the Gaudin eigenfunctions \citep{Gaudin}
\begin{equation}
\left|\left\{ E_{\nu}\right\} \right\rangle =\frac{1}{z}\prod_{\nu=1}^{n}\left(\sum_{j=1}^{N}\frac{A_{j}/2}{A_{j}/2+E_{\nu}}S_{j}^{+}+S_{0}^{+}\right)\left|\Downarrow\right\rangle .\label{eq:eigenstates}\end{equation}
that correspond to these eigenenergies. The unexcited state $\left|\Downarrow\right\rangle $
is all spins down, including the central spin, state. The normalization
factor is determinant of an $n\times n$ matrix \citep{Gaudin}, $z=\sqrt{\det\hat{M}}$,
which diagonal and the off-diagonal matrix elements are\begin{eqnarray}
M_{kk} & = & 1+\sum_{j=1}^{N}\frac{2l_{j}\left(A_{j}/2\right)^{2}}{\left(E_{k}+A_{j}/2\right)^{2}}-\sum_{p=1\neq k}^{n}\frac{2E_{p}}{\left(E_{k}-E_{p}\right)^{2}},\nonumber \\
M_{kk'} & = & \frac{2E_{k'}^{2}}{\left(E_{k}-E_{k'}\right)^{2}}.\label{eq:M_off_diagonal}\end{eqnarray}

This diagonlization procedure can be constructed in an easy way by
solving a complementary bosonic model instead of Eq. (\ref{eq:H0})
to obtain the ansatz for the eigenstate, Eq. (\ref{eq:eigenstates}).
Then, the Richardson equations emerge as the requirement for the states
in Eq. (\ref{eq:eigenstates}) to be the eigenstates of the model
Eq. (\ref{eq:H0}). In this way the exact form of Eq. (\ref{eq:Richardson_Eq})
can be found by using the spin commutation relations only. This approach
was developed in \cite{vonDelft} in the context of the BCS model
and in \cite{OTvonDelftLoss} in the context of the Dicke model.

When magnetic field is very large the last term in the Richardson
equations, Eq. (\ref{eq:Richardson_Eq}), can be neglected in leading
$1/\mu B$ order therefore all of the roots $E_{\nu}$ are close to
either $-A_{\nu}/2$ or $\mu B$. There is up to one root close to
$\mu B$ and, we consider here 1D case only, there are up to two roots,
$k_{j}\leq2$, close to $-A_{\nu}/2$. Collection of all possible
sets of $k_{j}$, such that $n=\sum_{j=1}^{N}k_{j}+\left(1\pm1\right)/2$
is equal to the number of excitations, explores the complete set of
the eigenstates in a given subspace of the Hamiltonian. We will use
$-A_{\nu}/2$ and $\mu B$ to label an eigenstate instead of $E_{\nu}$.

Corrections to these limiting values of the roots at a finite $B$,
$E_{\nu}=-A_{\nu}/2+\delta_{\nu}$ and $E_{\nu}=\mu B+\delta_{\nu}$,
can be found from the $1/\mu B$ expansion of the Richardson equations,
Eq. (\ref{eq:Richardson_Eq}), \begin{multline}
\delta_{\nu}=\left\{ \begin{array}{l}
-\frac{A_{\nu}^{2}}{2\mu B},\qquad\;\;\:\qquad\qquad\qquad\textrm{ if one \ensuremath{E_{\nu}}=\ensuremath{-A_{\nu}/2}},\\
-\frac{\left(1\pm i\right)A_{\nu}^{2}}{4\mu B},\qquad\qquad\qquad\quad\;\textrm{ if two \ensuremath{E_{\nu}}=\ensuremath{-A_{\nu}/2},}\\
\sum_{k=1}^{n-1}\left(A_{k}+\frac{\left(l_{k}-\frac{1}{2}\right)A_{k}^{2}}{\mu B}\right)\end{array}\right.\\
-\sum_{j=1}^{N}l_{j}\left(A_{j}-\frac{A_{j}^{2}}{2\mu B}\right),\textrm{ if \ensuremath{E_{\nu}=\mu B.}}\label{eq:delta_nu}\end{multline}
The sum $\sum_{k=1}^{n-1}$ in the third case is over the remaining
$n-1$ roots $-A_{\nu}/2$. Note that the corrections to the roots
$-A_{\nu}/2$ were obtained by linearizing the system of equations,
Eq. (\ref{eq:Richardson_Eq}), and the correction to the root $\mu B$
was obtained by expanding Eq. (\ref{eq:Richardson_Eq}) up to the
second order in $1/\mu B$ with the first order corrections to the
roots $-A_{\nu}/2$ obtained from the linearized equations.

The normalization factor $z$ can also be expanded in a $1/\mu B$
series and we find that the product of the diagonal matrix elements
of $\hat{M}$ gives in leading order \begin{equation}
z=\begin{cases}
\left(2\alpha\right)^{n}\prod_{k=1}^{n}\frac{1}{A_{k}}, & \textrm{ if all \ensuremath{E_{\nu}=-A_{\nu}/2,}}\\
\left(2\alpha\right)^{n-1}\prod_{k=1}^{n-1}\frac{1}{A_{k}}, & \textrm{ if one \ensuremath{E_{\nu}=\mu B}},\end{cases}\end{equation}
where the product in the second case is over the remaining $n-1$
roots $-A_{\nu}/2$. Upper bound of the first subleading correction
to $z$ can be estimated as $r_{0}^{2}\sqrt{n!}$, assuming that a
few off-diagonal matrix elements (\ref{eq:M_off_diagonal}) are of
the order of $r_{0}^{4}$ and they all contribute with the same sign
to the determinant, which is smaller than the leading terms if $\mu B\gg A_{0}\sqrt{n/e}$
when number of the excitations is large $n\gg1$ ($A_{0}$ is the
maximal coupling strength). We used the Stirling's formula to approximate
$^{n}\sqrt{n!}$ for a large $n$. At high external magnetic field,
$B\gg B_{\textrm{fluc}}$, the first subleading correction is small.

Performing summation over $E_{\nu}$ with the accuracy of Eq. (\ref{eq:delta_nu})
we obtain the eigenenergies of the Hamiltonian Eq. (\ref{eq:H0})
at high magnetic field, \begin{multline}
E=\pm\frac{\mu B}{2}\pm\sum_{j=1}^{N}\bigg[\frac{A_{j}\left(-l_{j}+k_{j}\right)}{2}\\
+\frac{A_{j}^{2}\left(2l_{j}-k_{j}+\frac{1\mp1}{2}\right)\left(k_{j}+\frac{1\pm1}{2}\right)}{4\mu B}\bigg],\label{eq:E_G}\end{multline}
where $\pm$ refers to the {}``up'' and {}``down'' electron Zeeman
levels and $k_{j}$ labels the number of roots of Eq. (\ref{eq:Richardson_Eq})
that are close to $-A_{\nu}/2$. Substituting $E_{\nu}$ with the
corrections from Eq. (\ref{eq:delta_nu}) into Eq. (\ref{eq:eigenstates})
we obtain the eigenfunctions that correspond to the eigenenergies
$E$, \begin{equation}
\left|\left\{ E_{\nu}\right\} \right\rangle =\left(S_{0}^{+}\right)^{\frac{1\pm1}{2}}\prod_{j=1}^{N}\left(S_{j}^{+}\right)^{k_{j}}\left|\Downarrow\right\rangle ,\label{eq:eigenfunctions_highB_G}\end{equation}
in leading $1/\mu B$ order. Corrections to this wave function contains
admixture of states from the opposite electron level with single nuclear
spin flip but each of them is proportional to a small factor $A_{j}/\mu B$.

The $1/B$-expansion of the Richardson equations in Eqs. (\ref{eq:E_G},
\ref{eq:eigenfunctions_highB_G}) coincides with the result in Eqs.
(\ref{eq:Ea},\ref{eq:eigenfunctions_highBa}) obtained using the $1/B$-expansion
of the main text.%
\begin{figure*}[!p]
\centering\includegraphics[width=0.9\textwidth]{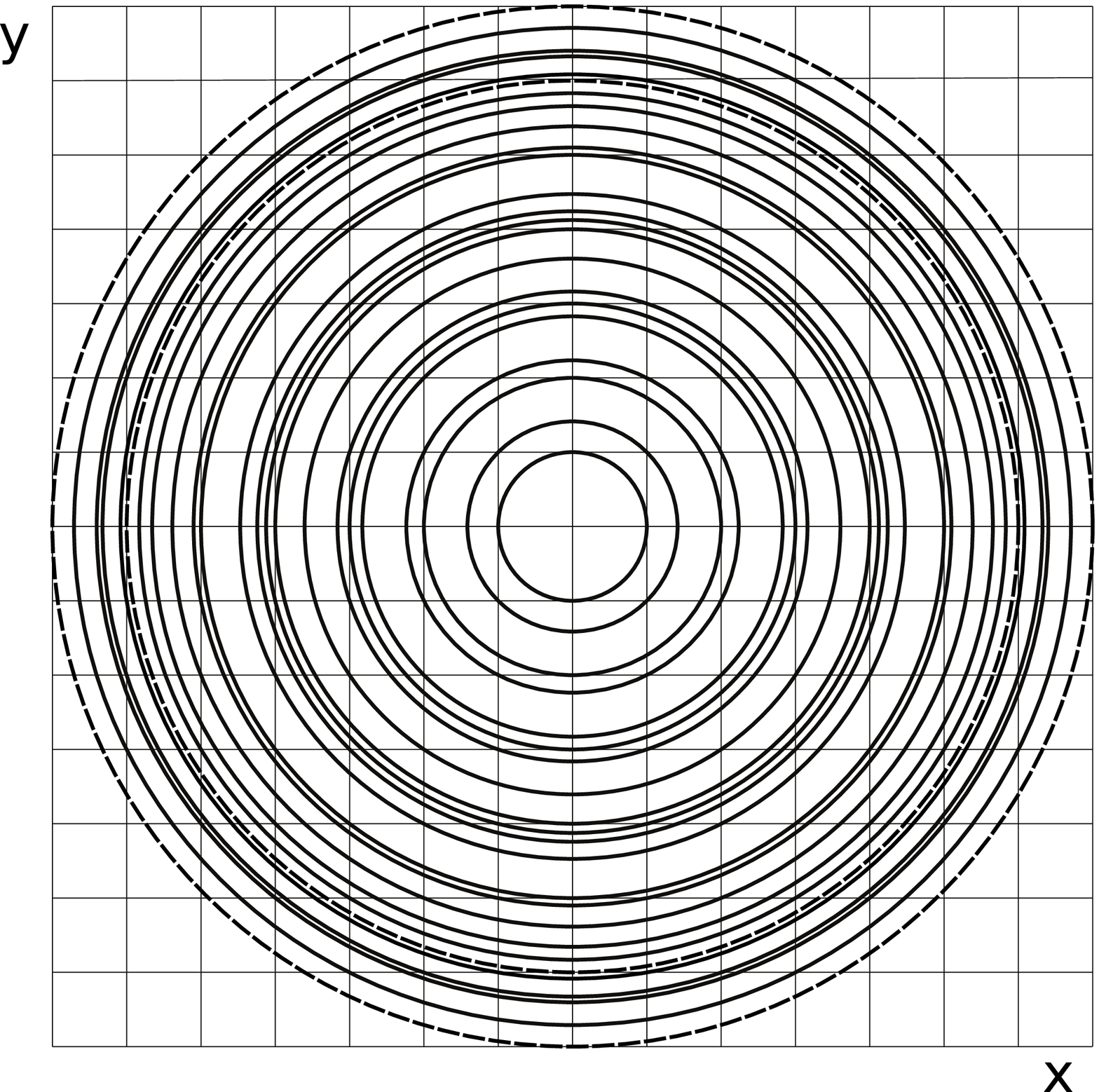}

\caption{Shells of constant couplings for a 2D square lattice. The majority
of the shells has degeneracy 8. Two circles marked by dashed lines
select a subgroup of shells between $r=5$ and $r=6$.}

\end{figure*}

\begin{figure*}[!p]
 \includegraphics[width=0.9\textwidth]{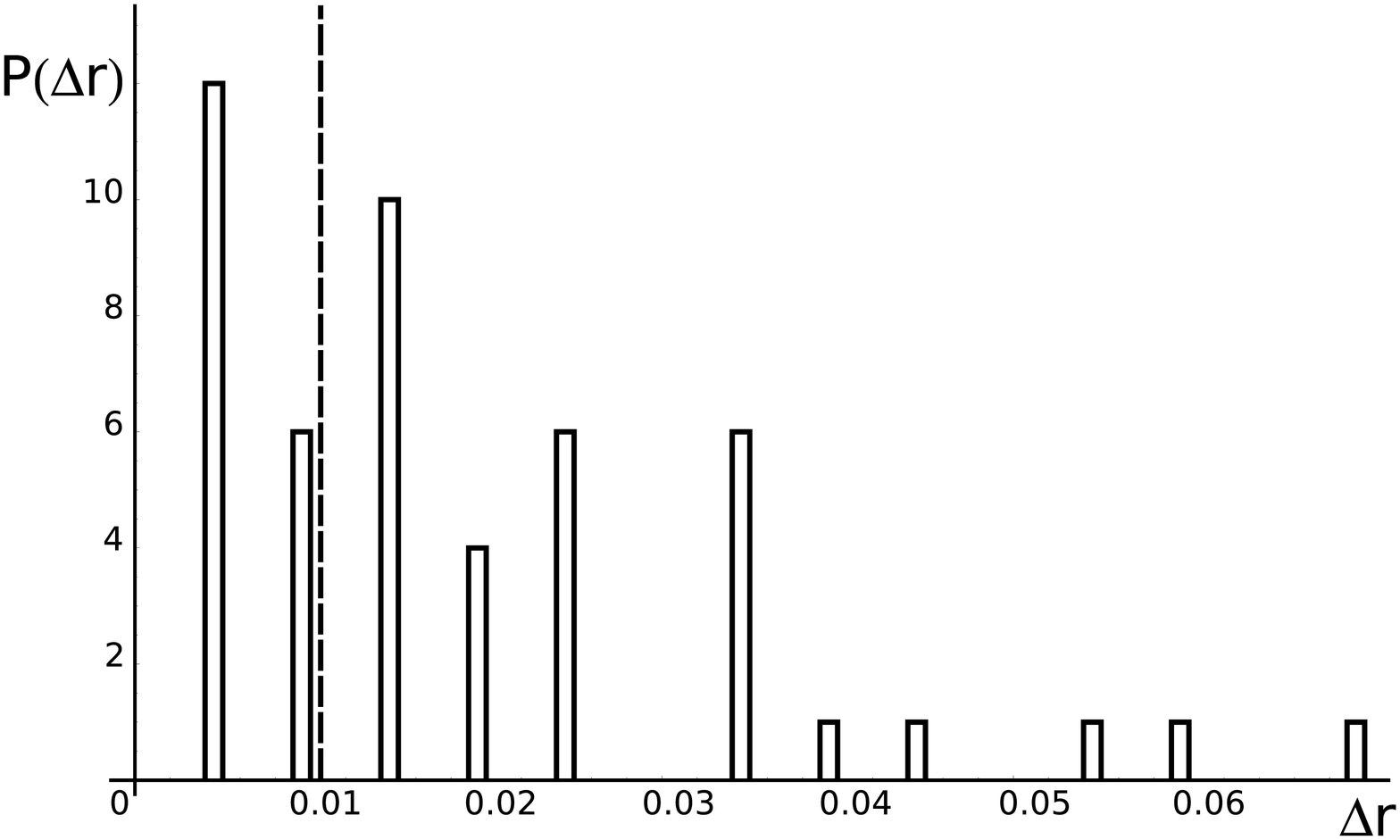}

\caption{Distribution of shell spacings for a 2D square lattice, $100<r<101$.
The dashed line marks the uniform spacing we use in the main text.}

\end{figure*}


\begin{thebibliography}{16}
\bibitem{degeneracy}Representation of $2S_{j}>2$ spins 1/2 using
a large angular momentum undercounts the number of the original states.
The undercounted states are degenerate in energy and have to be accounted
in a calculation of overlaps.
\end{thebibliography}
\end{document}